\documentclass[%
 reprint,
 amsmath,amssymb,
 aps,
]{revtex4-2}

\usepackage{graphicx}
\usepackage{dcolumn}
\usepackage{bm}
\usepackage[colorlinks=true, allcolors=blue]{hyperref}

\hypersetup{
    pdftitle={ORBIT},
    pdfauthor={William K. Black et al.}, 
    pdfsubject={BBH}, 
    pdfstartview={FitH}, 
    pdfkeywords={Dendro, Dendro-GR, supercomputing, black holes, binary inspirals, gravitational waves}
}

\usepackage[dvipsnames]{xcolor}

\definecolor{bleudefrance}{rgb}{0.19, 0.55, 0.91}
\definecolor{AmericanRed}{rgb}{0.698, 0.133, 0.204}
\definecolor{AmericanBlue}{rgb}{0.0391, 0.1914, 0.3789}
\definecolor{mustard}{rgb}{.808,.702,.00392}
\definecolor{ketchup}{rgb}{0.781, 0.160, 0.1328}
\definecolor{beeswax}{rgb}{.9137, .6235, .2471} 
\definecolor{PumpkinOrange}{rgb}{0.984, 0.49, 0.027}

\begin{document}


\title{Nyquist-resolving gravitational waves via orbital frequency-based refinement}

\author{\href{https://orcid.org/0000-0003-4811-7913}{William K. Black}}
  \email{wkblack@byu.edu}
\author{David Neilsen}
 \author{Eric W. Hirschmann}
\affiliation{
 Department of Physics and Astronomy, 
 Brigham Young University, \\
 N283 Carl F. Eyring Science Center, Provo, UT 84602
}

\author{David F. Van Komen}
\affiliation{
  Kahlert School of Computing, 
  University of Utah, 
  50 Central Campus Dr., Room 3190, Salt Lake City, UT 84112
}

\author{Milinda Fernando}
\affiliation{
    Oden Institute, 
    The University of Texas at Austin, \\
    201 E 24th St, Austin, TX 78712
}

\date{\today}

\begin{abstract}

Adaptive mesh refinement efficiently facilitates the computation of gravitational waveforms in numerical relativity. 
However, determining precisely when, where, and to what extent to refine 
when solving the Einstein equations poses challenges; 
several \textit{ad hoc} refinement criteria have been explored in the literature.
This work introduces an optimized resolution baseline derived \textit{in situ} from the inspiral trajectory (ORBIT). 
This method uses the binary's orbital frequency as a proxy for anticipated gravitational waves to dynamically refine the grid, satisfying the Nyquist frequency requirements on grid resolution up to a specified spin weighted spherical harmonic order. 
ORBIT sustains propagation of gravitational waves while avoiding the more costly alternative of maintaining high resolution across an entire simulation---both spatially and temporally. 
We find that enabling ORBIT decreases waveform noise by an order of magnitude 
and better resolves high-order wave amplitudes through merger. 
Combined with WAMR and other improvements, 
updates to Dendro-GR decrease waveform noise, 
decrease constraint violations, and boost refinement efficiency 
each by factors of $\mathcal{O}(100)$, 
while reducing computational cost by a factor of four. 
ORBIT and other recent improvements to Dendro-GR begin to prepare us 
for gravitational wave science with next-generation detectors. 

\end{abstract}

\maketitle


\section{Introduction} \label{sec:intro}

Mergers of black holes (BHs) and neutron stars in binary inspirals generate 
gravitational waves (GWs) detectable by the LIGO and Virgo GW observatories.
By matching observed signals to GW templates---precomputed for likely 
GW sources---we can estimate the likely progenitors of those GW events. 
The reliability of binary merger parameter estimation hinges on 
the accuracy, precision, and coverage of those templates. 

The next generation of GW detectors, including 
LISA~\cite{amaroseoane2017laserinterferometerspaceantenna}, 
Cosmic Explorer~\cite{reitze2019cosmicexploreruscontribution}, 
and the Einstein Telescope~\cite{Maggiore_2020}, 
are being planned for the next decade.  
These new detectors will be over an order of magnitude more accurate than current detectors (with higher signal-to-noise ratio), making them not only more sensitive to a wider range of GW sources, but also able to detect much longer portions of the inspiral signal.  Each of these advances in GW science places new demands on the extent and quality of the waveform libraries.  We need waveform templates that are at least an order of magnitude more accurate, for a larger class of possible binaries, and for longer portions of the inspiral~\cite{Purrer_Haster_20,Ferguson+21}.

The nonlinear equations of motion for gravity make calculating 
waveform templates for astrophysical sources challenging.  
While some methods approximate solutions for the early inspiral 
or late ringdown of binary mergers, the only consistent 
method to calculate the full waveform is to compute solutions 
of the full nonlinear equations with numerical relativity (NR) codes.  
Computing NR waveforms is challenging and expensive; 
even for the lower accuracy needed by today's detectors,
waveforms have only been computed for a relatively small part 
of the binary merger parameter space~\citep[see e.g.][]{Pompili:2023tna}. 
The task of computing waveforms of sufficient accuracy, duration, and variety 
(in mass ratio, spin, orbital eccentricities) 
to satisfy the requirements of next generation GW detectors 
is beyond the scope of current NR codes~\cite{Purrer_Haster_20, Ferguson+21}. 
Solving this problem will require novel computational approaches.




Numerical solutions of binary inspirals require adequate resolution 
on multiple scales, ranging from the size of the compact objects 
to their orbital scale to the wavezone to the outer boundary 
of the computational domain. 
In dynamical problems, Adaptive Mesh Refinement (AMR) 
efficiently handles these multiple scales by adjusting grid resolution
to concentrate computational resources close to the compact objects
and in regions of strong gravitational wave emission. 
This approach maintains high accuracy in critical regions and lower resolution in less dynamically important regions, optimizing computational efficiency. 
In practice, however, determining precisely where, when, and 
to what degree to refine can challenge and defy na{\"i}ve intuition.

Refinement strategies come with various levels of complexity 
and dependence on the solution of the equations solved. 
Broadly speaking, some refinement criteria use specific 
features known to occur in a solution, 
such as the location of a shock in hydrodynamics 
or a black hole's location in NR, 
while others rely on measures of the solution's convergence 
(which criteria can be solution independent).
As an example of the first type of refinement criteria,
NR codes can directly track black hole (BH) 
apparent horizons and puncture locations. Refinement can then be 
focused around these locations to ensure a certain resolution on 
the horizon and in its vicinity, structuring a nested grid of concentric 
spheres or boxes defined by empirically determined parameters 
\citep[see, e.g.][]{Schnetter+04, Zlochower+05, Campanelli+06, Radia+22, Rashti+24a}. 
Other proxies (e.g. functions of the conformal factor, 
the lapse, and the Hamiltonian and momentum constraints, and/or their derivatives) 
have also been widely used to determine refinement 
in binary inspiral simulations \citep[see, e.g.][]{Radia+22}.

Refinement criteria based on solution quality 
have also been widely used in numerical relativity.
Choptuik~\cite{ChoptuikPhD,ChoptuikFrontiers,Choptuik:1991bfu,ChoptuikApproaches}
was the first to use Berger-Oliger AMR~\cite{Berger_Oliger_1984} in NR, which 
estimates convergence using Richardson extrapolation to guide refinement.
Choptuik also advocated for a more efficient measure of self-convergence using a shadow hierarchy~\cite{1997ASPC..123..305C,Pretorius_2002}, 
which has been widely implemented (see~\cite{Thornburg:2010ick} for a partial bibliography).
Recently, \citet{Rashti+24a} monitored solution convergence using Lagrange interpolation. 
\citet{Radia+22} describes another method, monitoring truncation error between refinement levels as a criteria for additional refinement. 
Our approach to monitoring solution convergence uses a wavelet expansion
of the solution variables. The wavelet basis is generated by shifting and scaling
a mother wavelet, giving a compact basis set that naturally generates spatial refinement. 
Wavelet Adaptive Multi-Resolution (WAMR) expands in interpolating 
wavelets~\cite{Paolucci1,Paolucci2,DeBuhr+18}; 
we implement this method in Dendro-GR~\cite{Fernando+19a,Fernando+23}. 

A practical AMR scheme almost always needs 
a combination of the two broad approaches listed above.  
Refinement schemes based on a few specific features can miss other 
important aspects of the solution, while general schemes can waste resources 
by refining on physically unimportant regions or aspects of the solution. 
Thus, there always remains an element of intuition and experience in 
determining precisely how to parameterize each of these methods. 
This paper highlights our work using such a hybrid approach to refinement.

The current work introduces a novel refinement strategy which provides 
a resolution floor for GWs produced by compact binaries. 
Because the orbital frequency equals or supersedes the frequency 
of the emitted GWs (once rescaled appropriately), we can use it to create 
an \textit{in situ} criteria for resolving the GW Nyquist frequency. 
We dub this method ORBIT (see \S\ref{sec:ORBIT})
and implement it in the code Dendro-GR. 
We will show that this baseline resolution sustains GWs 
as they propagate through the grid, capturing full 
waveform amplitudes to a higher precision.

In this paper, we show how recent improvements to Dendro-GR (\S\ref{sec:Dendro}), 
including a refinement floor based on tracking inspiral frequency (ORBIT, \S\ref{sec:ORBIT}), 
permit us to simulate with improved accuracy, precision, and efficiency. 
\S\ref{sec:results} details these improvements, focusing on lower mass ratios $q = \{1, 4\}$; 
a future paper will catalog waveforms at larger mass ratios, $q \in [8, 32]$.

\section{Methods} \label{sec:Dendro}

The relativistic astrophysics code 
\href{https://github.com/paralab/Dendro-GR/}{Dendro-GR} 
\citep{DeBuhr+18, Fernando+17, Fernando+19a, 
Fernando+19b, Fernando+22b, Fernando+23} 
uses wavelet-based adaptive multiresolution (WAMR) to capture 
sharp features on an unstructured octree grid. 
Dendro-GR calculates wavelet coefficients for all 24 BSSN fields 
\footnote{
  We found that calculating wavelet coefficients for a subset 
  of the fields was insufficient to capture proper dynamics of mergers. 
  Scalar functions such as the conformal factor $\chi$ or lapse $\alpha$
  were too smooth to provide adequate resolution for all other fields. 
  Even after including $\tilde A$ and other non-scalar fields, it was 
  ultimately only when we included \textit{all} fields that we achieved 
  good behavior of our binary mergers
  \citep[c.f.][]{Radia+22}. 
} 
and triggers refinement 
if the wavelet expansion 
coefficients exceed a threshold magnitude $\epsilon(t; \vec x)$. 
Decreasing this threshold both increases the number of basis functions
used to represent the solution while also increasing grid resolution
in those regions which need it most, driving the solution toward
the convergent regime. 
An insufficiently strict wavelet tolerance could then 
preclude access to the convergent regime, 
failing to model crucial features of the evolution. 
WAMR responds to the resolution requirements of the wavelet basis, 
refining regions near the BHs and regions with strong GW signals 
while keeping regions far from the BHs at a lower grid resolution,
truncating the wavelet expansion. 
Dendro-GR scales well: in massively parallel jobs, 
work per processor remains roughly constant. 
This structure allows us to complete full inspirals 
in a matter of days to weeks for $q \lesssim 16$.

In this paper we use Dendro-GR with much of the same setup as in \citet{Fernando+23} 
(this code version is hereafter referred to as ``v2022''): 
utilizing the octree code Dendro for refinement, 
we generate initial data with {\sc TwoPunctures} 
\citep{Ansorg+04} and solve the BSSN equations. 

Unlike \citet{Fernando+23}, we make several alterations to the equations (\S\ref{sec:formalism}) and grid structure (\S\ref{sec:grid}). 
Our current formalism now incorporates 
  slow-start lapse \citep[][\S\ref{sec:SSL}]{Etienne_24} and customized Hamiltonian damping (HD; \S\ref{sec:HD}) into the BSSN equations. 
Three key strategies improve our grid structure:
  structural adjustments within the orbital radius (\S\ref{sec:onion}),
  causal wavelet adaptive mesh refinement (\S\ref{sec:cWAMR}), and 
  an evolving Nyquist refinement floor dictated by BH trajectories (ORBIT; see~\S\ref{sec:ORBIT}). 
The following subsections detail these enhancements to Dendro's core functionality. 


%

\vspace{-.5cm}
\subsection{Formalism} \label{sec:formalism} 

We solve the BSSN equations as in \citet{Fernando+23} 
but with two additions: 
  ``slow-start lapse'' (SSL) and 
  custom Hamiltonian damping (HD). 
Both work to reduce noise and error in the solution: 
  SSL at early times by decreasing noise generated by the initial lapse wave, 
  and HD at later times, damping Hamiltonian constraint violation over time.

\subsubsection{Slow-Start Lapse (SSL)} \label{sec:SSL}

A common challenge with the standard $1+\log$ slicing and 
$\Gamma$-driver shift conditions is the emergence of 
an initial gauge wave packet propagating outward from 
the BHs with superluminal speed $\sqrt{2} c$ \cite{Etienne+14}. 
If WAMR is equally applied across the space which the 
sharp gauge pulse propagates through, it will trigger
extensive refinement along the two-dimensional surface
associated with the outward-traveling wave. 
Dendro most efficiently resolves zero-dimensional 
(i.e. point-like) features, so resolving this 2D
surface incurs substantial computational cost. 
As this refinement occurs on an unphysical feature, 
it would be preferable to avoid such inefficient expenditures. 

Recent results from \citet{Etienne_24} show that adding 
a ``slow-start lapse'' (SSL; see \S\ref{sec:SSL}) condition 
to BSSN weakens the initial lapse wave. 
Particularly at high $q$, SSL spreads out the wave packet, 
reducing lapse wave sharpness and amplitude. 
This smoothing of the lapse wave decreases 
refinement necessary to resolve the peak 
and reduces constraint violations by 2--6 orders of magnitude. 
Together with other improvements, numerical noise in the waveform decreased by a factor of $\sim 4.3$, revealing previously obscured higher-order modes.

\citet{Etienne_24} defines SSL as adding the following term to the lapse evolution equation: 
\begin{equation}
  \partial_t \alpha = [\cdots] - W \left[ h \, {\rm e}^{- \frac{1}{2} t^2/\sigma^2} \right] (\alpha - W),
\end{equation}
where $W = \chi^{1/2}$ is the square root of the conformal factor (Dendro-GR evolves $\chi$) and the 
dimensionful constants $h$ and $\sigma$ are found via numerical experiment; 
in this work we use the default values of $h = 0.6 / M$ and $\sigma = 20\,M$. 
This additional term drives the initial lapse wave toward $\alpha = W$. 
SSL decreases the lapse wave's amplitude and frequency 
such that its Gaussian curvature (functional; not spacetime) 
is reduced by a factor of $\sim 8 \, q$; 
this both reduces noise on the grid and reduces refinement 
required by feature-dependent strategies like WAMR. 

We find that enabling SSL for a $q=1$ run reduces constraint violations 
by over an order of magnitude at early times and near merger 
decreases the necessary mesh size by a factor of three 
(due to reduced noise on the grid).  
Overall, this results in waveforms with roughly four times less noise. 
As discussed in \citet{Etienne_24}, SSL may shift merger time: 
we found for a low-resolution $q=1$ run a $\sim -4\,M$ shift 
(which decreased with higher resolution).

\subsubsection{Hamiltonian Damping (HD)} \label{sec:HD}

Our Hamiltonian damping scheme (HD) builds off previous work 
\citep{Yo_Baumgarte_Shapiro_2002, Duez_Marronetti_Shapiro_Baumgarte_2003, Etienne_24} 
which added the diffusive term $C \mathcal{H}$ 
(where $C$ is some dimensionful constant 
and $\cal H$ is the Hamiltonian constraint violation) 
to the evolution of the conformal exponent $\phi$. 
They found a Courant condition where $2C \Delta t / (\Delta x)^2 \leq 1$. 
We replace $C$ with $\frac14 c_H (\Delta x)^2 / \Delta t$
to account for dependence on local grid spacing and time step. 
As we evolve the conformal factor $\chi = {\rm e}^{-4 \phi}$ instead of $\phi$, we need to include an extra factor of $\chi$ in our diffusive term: 
\begin{equation}
  \partial_t \chi = [\cdots]  + c_H \chi \frac{(\Delta x)^2}{\Delta t} \mathcal{H} 
\end{equation}
(we use $c_H = 0.077$). 
This modification approaches zero in the the continuum limit. 

In our runs, HD rapidly lowers $\cal H$ by an order of magnitude, 
but these gains level off. 
The constant remeshing caused by the BHs traveling across the grid 
generates new $\cal H$ with each new interpolation of data 
(whether from refining or coarsening).  
Errors reach an equilibrium during inspiral, 
as the same amount of $\cal H$ generated at each remesh is removed 
by the next remesh step, sustaining a somewhat steady $\cal H$ floor. 
As with SSL, HD causes a slight shift in merger time 
($\sim +1.5\,M$ for a low-resolution $q=1$ run, 
decreasing with higher resolution).

\subsubsection*{}

We have found that the two modifications above (SSL + HD) 
allow Dendro-GR to run at about half the computational cost 
while reducing constraint violations by over a factor of twenty.

\subsection{Grid Structure} \label{sec:grid} 

Dendro-GR is built atop the computational framework Dendro, which uses an octree to represent the computational domain as an adaptive unstructured grid. 
Dendro subdivides the initial computational domain into octants dependent on features present in the initial data through a wavelet-based approach, iterating until refinement is no longer required for accurate computation, or until a maximum refinement level is reached. 
Dendro subjects all octants to a 2:1 grid balancing condition---this ensures 
that any given octant differs at most by a single level from its neighbors. 
\citet{Fernando+23} provides an in-depth explanation of the initial grid generation and refinement procedures.

Our modifications require the grid structure to include hard-coded elements depending on current BH locations (Onion; see \S\ref{sec:onion}), the current time of the simulation (causal WAMR; see \S\ref{sec:cWAMR}), and current BH velocities (ORBIT; see \S\ref{sec:ORBIT}). 

%

\subsubsection{Onion: recursive inner grid structure} \label{sec:onion}

Different parts of a grid call for different refinement strategies. 
We must ensure sufficient resolution while striving for an efficient implementation.  
The BHs in the central region of the grid each require sufficient 
resolution throughout the simulation to accurately generate waveforms. 
We therefore enforce a base, core refinement around each BH. 
However, we cannot sustain such a high level of resolution across the entire 
simulation---we must relax that refinement as we recede from the BHs; 
the rate at which that derefinement occurs can affect the waveform. 
Furthermore, at the orbital scale we find that we cannot have reverted to too little refinement. 
We discuss our refinement approaches at each of these scales. 


\paragraph{Core refinement} 

In the vicinity of each BH, we manually set the refinement level 
to ensure a minimum of 50 grid points across each (nonspinning) 
horizon (see Appendix~\ref{apx:central_refinement}). 
This core, maximum refinement level $\ell_{{\rm BH}, i}$ 
encapsulates each BH to a radius of $R_{{\rm AMR}, i} \sim 2 \, m_i$.

\paragraph{Neighboring refinement}


As mentioned above, Dendro's octree structure has a 2:1 derefinement criteria, which requires that cell neighbors differ at most by a single refinement level \citep{Sundar+08, Fernando+23}. 
While this prevents \textit{immediate} derefinement by more than one level, 
physically proximal regions could still drastically differ in resolution. 
Because coarser grids have a higher effective impedance for signals (acting like stiff barriers), high-frequency signals can suffer significant back-reflection on encountering refinement boundaries of lower resolution regions \citep{Radia+22}.  
Without gradual derefinement about the BHs, the outward propagating initial lapse wave will tend to immediately back-reflect onto the BHs, generating large constraint violations and potentially causing an unphysical influence on the BH trajectories. 
To help alleviate these rapid shifts in grid spacing, we construct buffer regions, further limiting the rate at which the grid coarsens beyond the 2:1 condition. 
In \citet{Radia+22}, the authors implement buffer regions by demanding a minimum number of cells between consecutive refinement level boundaries, thus spatially slowing derefinement. 
We elect instead to use a physical criteria for grid derefinement, 
dependent on the distance from each BH. 

As we move away from the highly resolved BHs, 
we only permit the refinement level $\ell$ to decrement 
once the radial distance from the BHs has increased 
by a geometric factor $r_{\ell-1} \sim \gamma \, r_{\ell}$. 
Explicitly, the resolution floor follows
\begin{equation}
  \ell \geq \ell_{{\rm BH}, i} - \log_\gamma \max\left( 1, \, r_i / R_{{\rm AMR}, i} \right), 
\end{equation}
where again $\ell_{{\rm BH}, i}$ defines the core refinement level within radius $R_{{\rm AMR}, i}$ about each BH; 
the geometric ratio $\gamma$ then relaxes between levels. 
We need a factor of $\gamma > 1$ in order for consecutive radii 
to expand and provide a buffer region to reduce back-reflections. 
If we were to use a factor of $\gamma > 2$ then the radii expand faster than the octree 2:1 grid structure, and BH positions could influence refinement in very distant, spacelike separated regions of the grid. 
We choose the golden ratio ($\gamma \approx 1.618$) as a number between these bounds; this marginally improves grid efficiency over $\gamma = 2$. 
We find that this gradual derefinement suppresses spikes in constraint violations (lowering total constraint violation by $\sim 4 \times$ for a low-resolution $q=1$ run), 
buffering the initial solution against early, high-amplitude echoes of the lapse wave on the BHs, reducing net constraint violation.

\paragraph{Orbital scale}

At the orbital scale, we enforce a wide base level of refinement to 
minimize changes in grid structure induced by changes in BH locations. 
As mentioned in \S\ref{sec:HD}, each change to grid resolution 
generates Hamiltonian constraint violations $\cal H$; 
we therefore desire to minimize changes to grid structure. 
To this end, we keep all points within a certain coordinate radius 
at a higher refinement level, $\ell_{\rm orbit}$. 
We set this coordinate radius at the orbital scale to be 
\begin{equation}
  R_{\rm orbit} = \max(r_1, r_2) + B,
\end{equation}
where $r_i$ is the current radial distance of the $i^{\rm th}$ BH from the grid center and $B$ is a buffer length. 
For our particular grids and grid sizes, we use $\ell_{\rm orbit} = 9$ (which corresponds to a $\Delta x \approx 0.26\,M$), which gives roughly a tenth the resolution of the BHs themselves. 
The buffer length permits the high refinement immediately about each BH to derefine to $\ell_{\rm orbit}$; for the scales in our simulations, we use $B = 8\,M$. 
In addition to minimizing grid level changes within $R_{\rm orbit}$, 
this base level of refinement also keeps the outer regions 
of the grid from changing structure with BH rotation. 
(As an example of this, in the absence of this floor, if the BHs were at angles $45^\circ$ and $225^\circ$ in the $xy$-plane, then Dendro would instruct the octants at $135^\circ$ and $315^\circ$ to derefine---then re-refine after the BHs moved another $45^\circ$.) 
This base level of refinement thus limits the changes to refinement level in both the inner regions and outer regions of the grid as the BHs orbit the origin. 
We find that enabling this orbital-scale refinement floor reduces Hamiltonian constraint violation $\cal H$ by about a factor of two compared to runs without it.

\begin{figure*}[!ht]\centering
  \includegraphics[width=\linewidth]{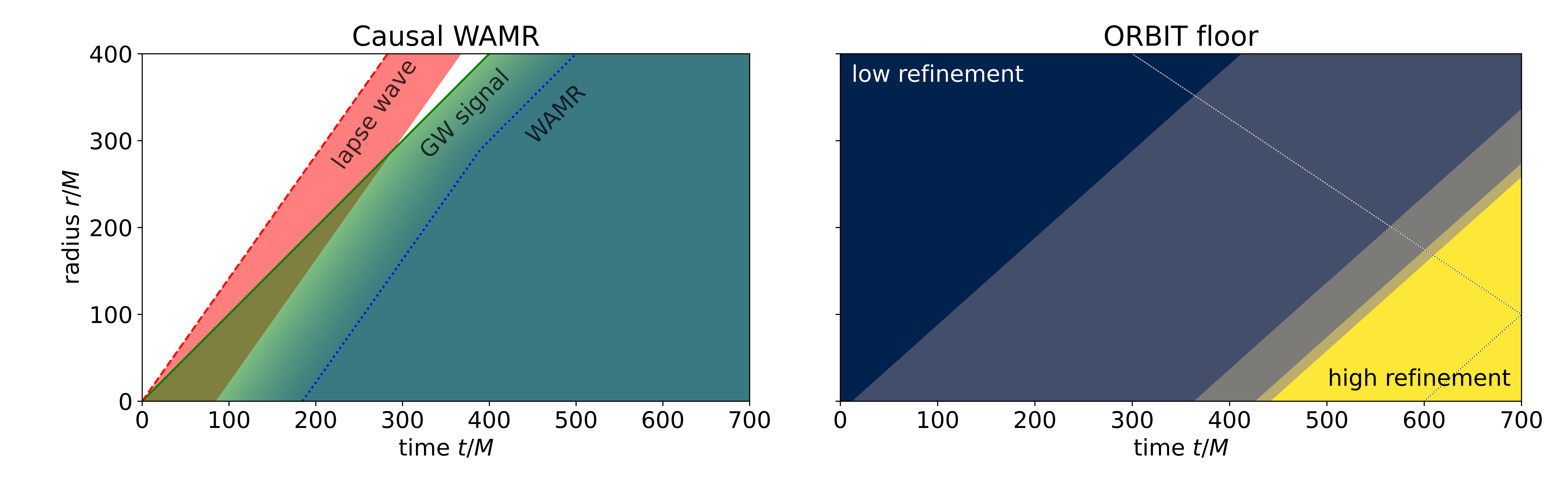}
  \vspace{-.75cm}
  \caption{
    Cartoon illustrating two novel refinement strategies recently implemented in Dendro-GR. 
    \textbf{Left:} Causal WAMR, a new refinement function dictating wavelet tolerances $\epsilon$ across the grid. We only activate WAMR in the causally-connected and clean regions of the grid, ignoring the lapse wave (red) and waiting for the GW signal (green) to fill the grid with physical data before refining. After these two events propagate outward, we gradually lower $\epsilon$ to its target value in a given region, requiring strict refinement in the blue-tinted region. 
    \textbf{Right:} ORBIT, a new refinement floor based on orbital frequency; lighter colors indicate higher refinement levels. 
    We only activate ORBIT in regions causally connected to the event of the gravitational wave ringdown striking the largest extraction radius (indicated by dotted white \& black lines; see Eq.~\ref{eqn:causal_end}). 
  }
  \label{fig:progress_grid}
\end{figure*}

\subsubsection{Causal WAMR refinement} \label{sec:cWAMR}
The initial lapse wave (see \S\ref{sec:SSL}) posed a significant challenge in our previous paper \citep{Fernando+23}. 
The 2022 version of the code used wavelets across the entire grid, 
keeping the orbital region highly refined with wavelets from the start of the simulation. 
This invoked a large computational cost at high mass ratios: 
the initial sharp lapse pulses propagated outward from the BHs in spherical shells, 
demanding refinement on their two-dimensional surfaces 
(while Dendro most efficiently resolves point-like, zero-dimensional objects). 
Enabling WAMR globally at simulation start caused a steep mountain of 
initial refinement as the lapse wave dispersed. 

While SSL ameliorates the issue by smoothing the lapse wave, it does not eliminate the issue entirely. 
We use a new refinement function which only activates refinement at times causally connected to the BHs from the initial time onward. 
We additionally avoid the noisy wake of the lapse wave. 
These two constraints activate WAMR only at times 
\begin{equation}
  t > \max \left( r, \frac{r + L_\alpha}{\sqrt{2}} \right) 
\end{equation}
where $t$ is the current time of the simulation, $r$ is the radial 
coordinate, and $L_\alpha = 120\,M$ is the width of the lapse wave. 
We linearly fade in over a time $t_{\rm fade} = 100\,M$ from a wavelet tolerance of $\log_{10} \epsilon = -3$ (effectively disabling wavelet refinement) to a runtime value of $\log_{10} \epsilon = -5$. 

This refinement strategy enables WAMR in only clean, causally-connected regions of the grid, disabling it in regions of the grid spacelike separated to the BH evolution and regions influenced by the lapse wave's propagation (see Figure~\ref{fig:progress_grid}). 
By avoiding causally disconnected and noisy regions, we can increase the sensitivity of our wavelet tolerance $\epsilon$ across the grid, thereby lowering constraint violations and generating more accurate waveforms. 

In addition to the gradual activation of WAMR, we spatially decrease 
wavelet tolerance $\epsilon$ with respect to radius. We use a 
constant value of $10^{-5}$ within the orbital radius $r \leq 8\,M$ 
and the value of $10^{-3}$ in outer regions $r \geq 400\,M$. We linearly
interpolate $\log \epsilon$ with respect to $\log r$ between those boundaries.

\subsubsection{ORBIT resolution floor} \label{sec:ORBIT}

The Nyquist frequency \citep{Shannon_1949} determines the maximum frequency (minimum wavelength) one can measure on a grid. If there are fewer than two points per wavelength of a signal, it will not transmit properly, distorted by aliasing. 
In audio processing, 1.1 to 5 times the Nyquist frequency is often used as a sampling rate to provide ``lossless'' signals \footnote{
  The upper range of human hearing is $\lesssim 20~{\rm Hz}$; CDs sample at 44.1~kHz or 48~kHz, about 1.1 to 1.2$\times$ the Nyquist frequency (40~kHz). 
  \href{https://www.ni.com/en/shop/data-acquisition/measurement-fundamentals/analog-fundamentals/acquiring-an-analog-signal--bandwidth--nyquist-sampling-theorem-.html}{National instruments} suggests sampling at 1.5--2.5$\times$ the Nyquist frequency \citep{NI_Signal}. 
  \href{https://robotae.com/newsroom/are-you-abusing-nyquist/}{Robotae} suggests $5 f_{\rm Nyquist}$ as a rule of thumb \citep{Rathbone_2021}. 
}. 
Beyond the Nyquist frequency, error goes down with sampling frequency squared---sampling at twice the Nyquist frequency decreases errors by about a half decade compared to sampling \textit{at} the Nyquist frequency (see \S\ref{apx:error_Nyquist}). 

We measure GWs with spin weighted spherical harmonics $(l,m)$ of the Weyl scalar $\Psi_4$ \citep{Newman_Penrose_1962}. 
Properly transmitting GWs demands grid resolution $\Delta x \leq \lambda_{\rm Nyquist} = \lambda_{\rm GW} / 2$. 
Because we will not always know the precise GW structure of a simulation \textit{a priori}, we cannot easily use the GWs themselves to make the grid Nyquist compliant. 
During the early inspiral, the orbital frequency of the black holes matches the gravitational wave frequency.  
In particular, the orbital frequency $f_{\rm orbit}$ will match the GW frequencies rescaled by their order $m$: $f_{\rm GW}^{(l,m)} / m$. 
Approaching merger, the black holes' orbital frequency outpaces that of the GWs they emit (see e.g. Figure~\ref{fig:omega-lambda}) and the BHs pass into a common horizon; beyond this point, their coordinate positions causally disconnect from the rest of the solution. 
Because $f_{\rm orbit} \geq f_{\rm GW}^{(l,m)} / m$, rescaling the BH orbital frequency to some desired mode $m$ provides a sufficient standard for estimating GW frequencies. 
We may thus satisfy the Nyquist criterion by tracking BH orbital frequency, calculating the corresponding refinement requirement, and propagating that requirement outward through the grid at the speed of light (matching the outward propagating GWs). 

In summary: to ensure Nyquist resolution of a gravitational wave mode $m_{\max}$, it is sufficient to mandate at radius $r$ and time $t$ that the grid spacing $\Delta x(t,r)$ relate to retarded orbital frequency $f_{\rm orbit}(t-r)$ as follows: 
\begin{equation} \label{eqn:dx_ORBIT}
  \Delta x(t,r) \leq \frac{c}{2 \, m_{\max} f_{\rm orbit}(t-r)} 
  \leq \lambda_{\rm Nyquist}^{(l,m_{\max})}(t,r).
\end{equation}
While this prevents \textit{total} back-reflection of low frequencies $f < m_{\max} f_{\rm orbit}$, partial back-reflection will still occur to some degree (with error following Eq.~\ref{eqn:error_Nyquist}). 
To mitigate back-reflections, one should pick $m_{\rm max}$ somewhat larger than the target spin weighted spherical harmonic order. 
In this paper, we use twice the frequency, Nyquist-resolving $m_{\rm max} = 12$ to clean $m = 6$. 
We dub this refinement scheme \textbf{ORBIT}: 
an optimized resolution based on the inspiral trajectory. 

Early in the simulation, low orbital frequencies require minimal refinement, but at plunge orbital frequencies rapidly increase (and then flatline; see Figure~\ref{fig:omega-lambda}), resulting in a stricter refinement criterion post-merger. 
This high refinement post-merger need only be used late in the simulation, and only in past timelike regions relative to the event where the last clean signal reaches the outermost radius at which GWs are measured $R_{\rm GW, \max}$. 
In particular, we disable ORBIT $\sim 122 \, M$ after merger at the largest gravitational wave extraction radius $R_{\rm GW, \max} = 100\,M$ and in all points not causally influencing that final point. 
That is: for all regions 
\begin{equation} \label{eqn:causal_end}
  |r - R_{\rm GW, \max}| > (t - t_{\rm end}),
\end{equation}
using $t_{\rm end} = t_{\rm merge} + R_{\rm GW, \max} + t_{\rm ring}$, with ringdown duration $t_{\rm ring} = \{22, 72\}\,M$ past merger for mass ratios $q = \{1, 4\}$ \footnote{
  The delay of $22\,M$ past merger comes from an analysis of ringdown modes in \citet{Giesler+19}: Analyzing data displayed in their Figure~8, we see at approximately $22\,M$ past merger the primary mode becomes $10$ times larger than the secondary mode. Perturbation theory can then take over, modeling waveforms from there on out. 
}. 
ORBIT therefore gradually increases the refinement of the grid from the inside out as we approach merger and dials down refinement in the outer and inner regions post-merger. 
As compared to retaining post-merger resolution globally, ORBIT's causal activation and deactivation reduces mesh size by an order of magnitude, 
improving our ability to place refinement only where (and when) it is needed.

\subsection*{}


Figure~\ref{fig:progress_grid} illustrates our two novel refinement strategies: causal WAMR and ORBIT. 
The left plot shows how causal WAMR acts on a clean and data-filled grid, in the future lightcone of the simulation. 
The right plot shows how ORBIT requires a resolution level dictated by the orbital frequency. 
The two refinement strategies combine to clean waveform noise by an order of magnitude compared to our previous work.

\section{Results} \label{sec:results}

In this section, we present improvements achieved by enabling ORBIT and other adjustments as we upgrade Dendro-GR from its 2022 version (``v2022'') to its 2025 version (``v2025''). 

In this paper, we focus on mass ratios $q=1$ and $q=4$. 
Mass ratio $q=1$ is of particular interest for ORBIT, as $q=1$ has the fastest orbital frequency (of non-spinning, non-eccentric mergers) and therefore stricter refinement constraints than higher mass ratios. 
We include a mass ratio $q=4$ run, leaning toward larger mass ratios.  
We have not reduced eccentricity for either of the mass ratios; this allows for one-to-one comparison to v2022.
Future work will reduce eccentricities and delve into $q \geq 8$ in more detail. 

For mass ratios $q = \{1, 4\}$, we compare enabling ORBIT (``ORBIT on''; $m_{\max} = 12$, aiming to resolve well $m=6$) to disabling ORBIT (``ORBIT off''; $m_{\max} = 0$). 
For v2022, only $q=4$ is available for comparisons (without ORBIT), and then only spherical harmonic modes $|m| \leq l \leq 4$. 

We first show that ORBIT enhances waveform precision: reducing noise and better propagating higher-order modes (\S\ref{sec:waveform_fidelity}). 
We also show how v2025 reduces constraint violations and improves efficiency of grid structure (\S\ref{sec:constraints}). 
Finally, we compare run times between v2022 and v2025, illustrating reductions in both wall time and total cpu hours (\S\ref{sec:run_times}).

\subsection{ORBIT enhances waveform fidelity} \label{sec:waveform_fidelity}

\begin{figure*}\centering
  \includegraphics[width=.49\linewidth]{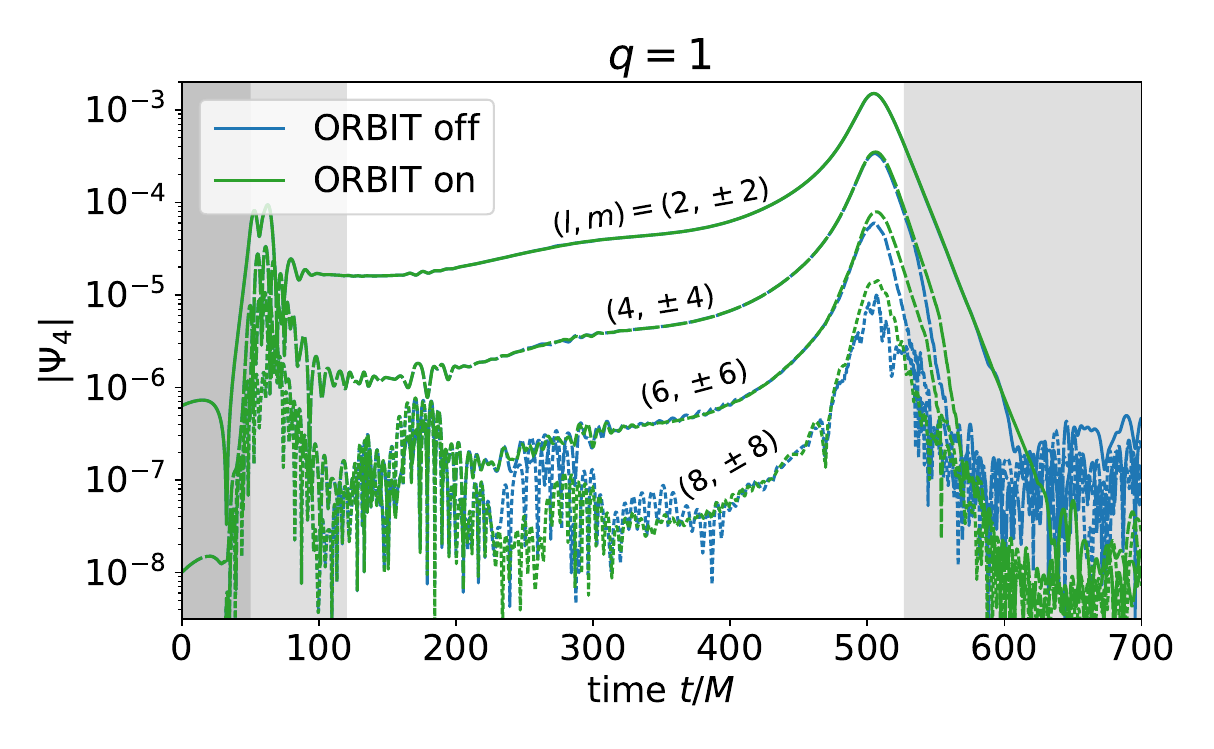}
  \includegraphics[width=.49\linewidth]{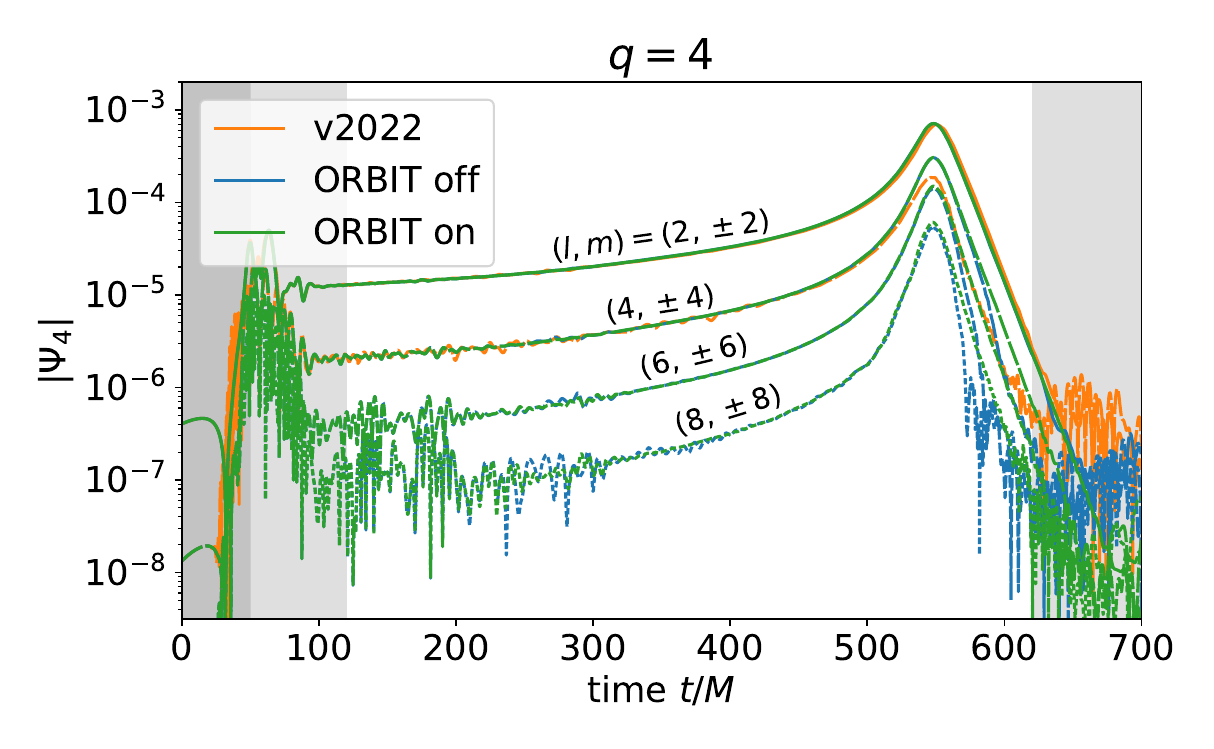}
  \vspace{-.5cm}
  \caption{
    Waveform amplitudes (measured at $R_{\rm GW} = 50\,M$) with and without ORBIT enabled ($m_{\rm max} = 12$ to enable; 0 to disable); 
    the end time of the right plot shows at a glance that recent improvements lower the noise floor by two decades. 
    Left and right plots give Weyl scalar $\Psi_4^{(l,m)}$ spherical harmonic amplitudes for mass ratios $q=1$ and $q=4$ respectively. 
    We here show modes $(l,m) = (x,-x)$ for $x \in \{2, 4, 6, 8\}$ in order from top to bottom. 
    The right plot additionally includes results from the 2022 version of Dendro-GR (orange; ``v2022'') for comparison to the new 2025 version of Dendro-GR (blue, green); this only has modes $x \in \{2, 4\}$ available for comparison. 
    Shaded regions on the left of each plot indicate the causally disconnected regime ($t < R_{\rm GW}$) and the region corrupted by the initial lapse wave ($t/M < (L_\alpha + R_{\rm GW})/\sqrt{2} \doteq 120.2\,M$). 
    Shaded regions on the right of each plot indicate where we terminate ORBIT support. 
    Note that none of the simulations were eccentricity-reduced; the $q=1$ run has large eccentricity modulating the amplitude. 
  }
  \label{fig:Psi4_qx}
\end{figure*}

\begin{figure}\centering
  \includegraphics[width=\linewidth]{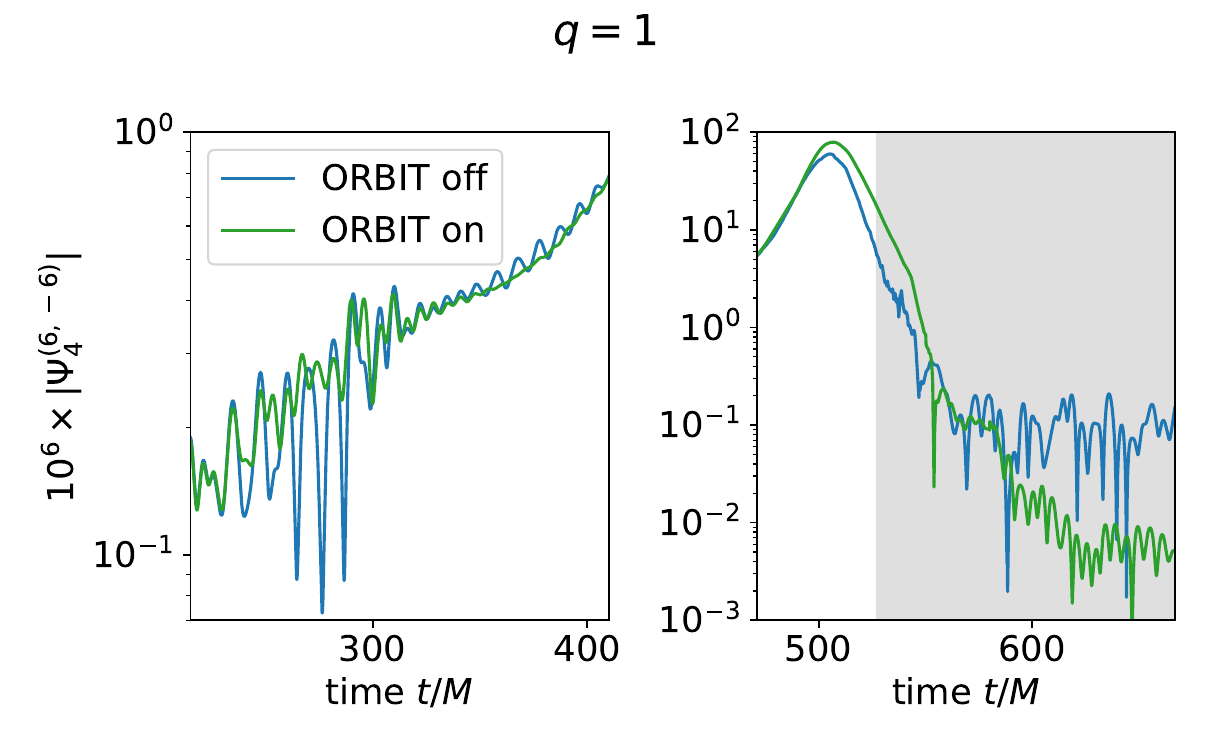}
  \vspace{-.75cm}
  \caption{
    Zoom-in on several notable regions of the left plot of Figure~\ref{fig:Psi4_qx} ($q=1$; identical coloring), focusing on the $(l,m) = (6,-6)$ mode. 
    ORBIT improves waveform quality both in the inspiral (left) as well as at merger and ringdown (right). 
  }
  \label{fig:Psi4_l6m6_cf}
\end{figure}

\begin{figure}\centering
  \includegraphics[width=\linewidth]{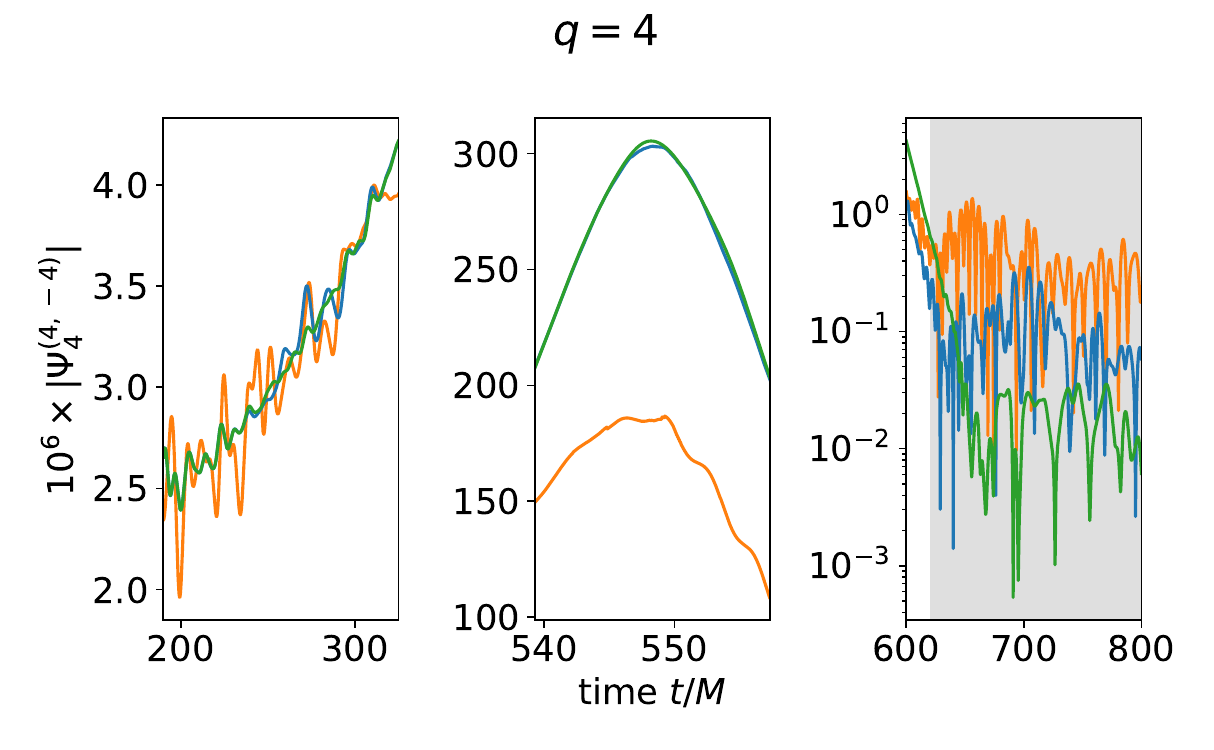}
  \vspace{-.75cm}
  \caption{
    Zoom-in on several notable regions of the right plot of Figure~\ref{fig:Psi4_qx} ($q=4$; identical coloring), focusing on the $(l,m) = (4, -4)$ mode. 
    Compared to the v2022 (the old code version), v2025 reduces noise in inspiral waveform (left), better captures peak waveform at merger (middle), and lowers the noise floor by nearly two orders of magnitude (right). 
  }
  \label{fig:amp_q4_zoom}
\end{figure}


In this section, we showcase the improved fidelity of our measured gravitational waveforms, focusing in particular on the effects of enabling ORBIT, our Nyquist-compliant orbital-frequency-based resolution floor. 
Figure~\ref{fig:Psi4_qx} displays Weyl scalar $\Psi_4$ waveform amplitudes for $q=1$ and $q=4$ mass ratios for ORBIT disabled (blue; ``ORBIT off''), for ORBIT enabled (green; ``ORBIT on''), and for the previous code version (orange; ``v2022''; ORBIT unavailable). 
Both left and right panels show spin weighted spherical harmonic modes $(l,m) = (x, -x)$ with $x \in \{2, 4, 6, 8\}$ (though v2022 only has $l \leq 4$ available). 
Shaded regions approximately indicate portions of the signal made noisy by the initial conditions (early times) and regions where perturbation theory can take over from numerical relativity (late times).

We see a clear reduction in waveform noise on enabling ORBIT, not only in the ringdown (where ORBIT demands its highest refinement), but also in the early inspiral (where ORBIT makes minimal demands). 
Noise reduction occurs in all modes but is most apparent at higher orders. 
Focusing on the $(l,m) = (6,-6)$ mode (the target of ORBIT's $m_{\max}$ criterion); Figure~\ref{fig:Psi4_l6m6_cf} shows roughly an order of magnitude reduction in waveform noise in the early inspiral for $q=1$ (we see about half an order of magnitude reduction for $q=4$). 


After plunge, ORBIT ensures that the proper peak amplitude of high-order modes propagates outward. 
Disabling ORBIT results in a $24\%$ reduction in merger amplitude for the $(6,-6)$ mode for $q=1$. 
For $q=4$, the same mode has a higher amplitude, and disabling ORBIT results in only a $6\%$ reduction. 
Without ORBIT's added resolution floor, higher order modes (which have lower amplitudes) fail to propagate through the grid to the GW extraction radii. 
After transmitting peak amplitude, ORBIT continues, carrying the ringdown outward. 
WAMR (our primary refinement scheme) ignores low-amplitude signals, so the exponential decay of ringdown amplitude is lost without ORBIT's sustaining influence. 

Figure~\ref{fig:amp_q4_zoom} highlights waveform improvements gained on upgrading the code from v2022 to v2025. 
For the $(l,m) = (4, -4)$ mode of $q=4$, we see roughly a factor of four reduction in waveform noise in the early inspiral (left panel). 
At merger (center panel), v2025 better captures peak waveform amplitude, matching trends discussed in appendix~\ref{apx:merger_amp}. 
Post-ringdown, the noise floor (right panel) drops by roughly a factor of ten on moving from v2022 to ORBIT off v2025, then by an additional factor of ten on enabling ORBIT. 
As next-generation detectors lower the noise floor on GW observations, the systematic errors of NR simulations must drop a commensurate amount. This shows progress in that direction. 

Discussed more in appendix~\ref{apx:merger_amp}, mode coupling strengthens toward more extreme mass ratios, increasing the relative importance of higher-order modes. 
Figure~\ref{fig:Psi4_qx} exemplifies this in the decreased distance between each of the $q=4$ modes (less than an order of magnitude between modes) compared to $q=1$ (over an order of magnitude between modes during inspiral). 
The trend with mass ratio continues, such that higher-frequency modes are of greater importance (to reconstruct observed waveforms) for higher mass ratios than they are for lower mass ratios. 
We must therefore resolve higher-frequency modes well for asymmetric mass ratios if we are to faithfully reconstruct the outgoing waveform.

%

\subsection{Code updates efficiently reduce constraints} \label{sec:constraints}

Updating our code from v2022 to v2025 both improves waveform quality and reduces net constraint violations. While globally refining the v2022 run would certainly reduce constraint violations, we quantify how well our improvements outpace the expected improvements of adding refinement to a 6th order code like ours, showing that our improved strategies refine more judiciously than before. 

\begin{figure}\centering
  \includegraphics[width=\linewidth]{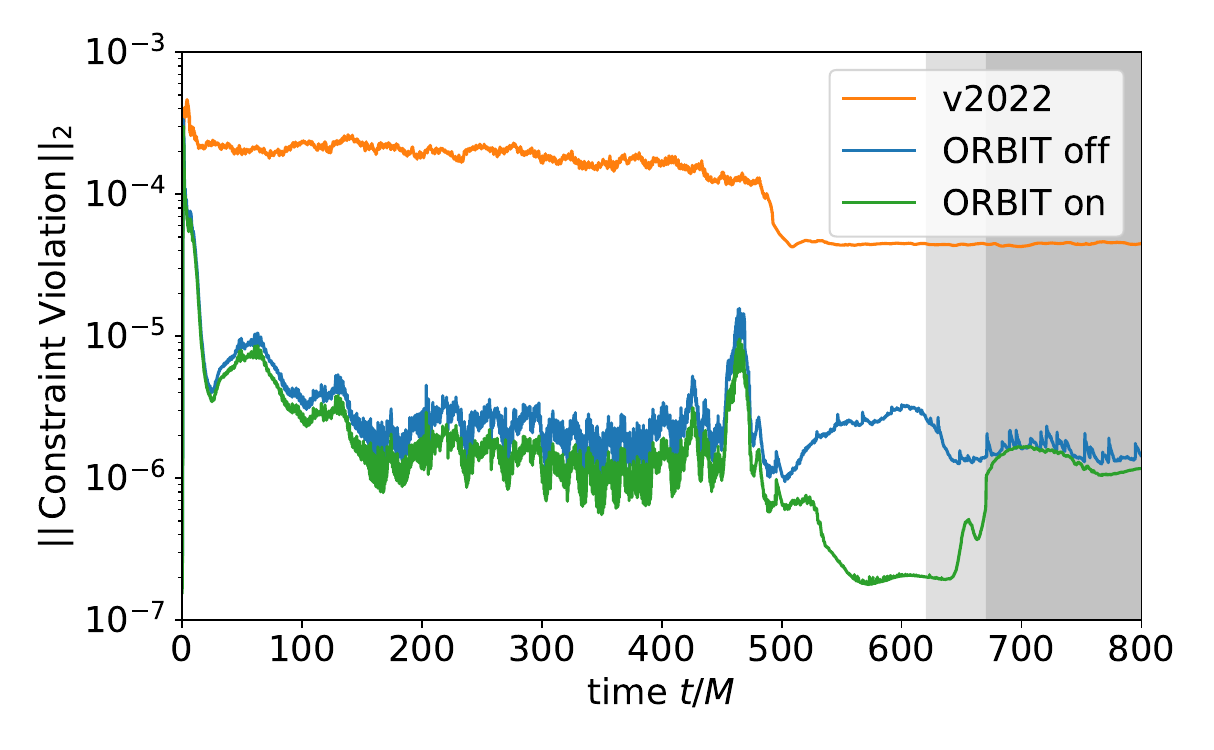}
  \includegraphics[width=\linewidth]{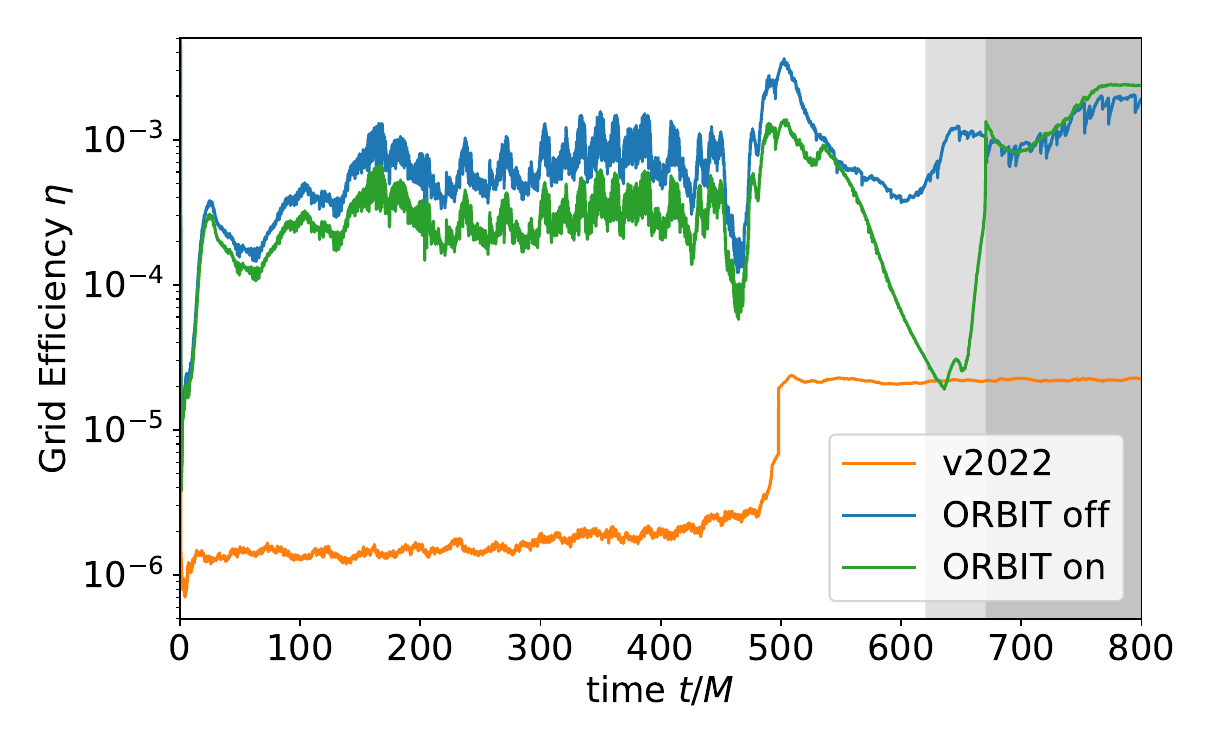}
  \vspace{-.75cm}
  \caption{
    Comparison of total constraint violation (upper panel; L2 norm of $\cal H$ \& $\mathcal{M}_i$ across grid) \& grid efficiency (lower; see Eq.~\ref{eqn:eta}) between code versions for $q=4$. 
    We see 10--100$\times$ lower constraint violations as we upgrade from v2022 to v2025. These improvements are executed efficiently, outperforming by $100\times$ the expected gains from globally refining a 6th order code. 
    Coloring as in Figure~\ref{fig:Psi4_qx}; shading indicates ORBIT disabling at gravitational extraction radii of $R_{\rm GW} = \{50, 100\}\,M$. 
  }
  \label{fig:CM}
\end{figure}

The upper panel of Figure~\ref{fig:CM} shows reductions in constraint violations between code versions for $q=4$ in moving from v2022 to v2025. 
Net constraint violation decreases by two orders of magnitude  
while Hamiltonian constraint decreases by up to three orders of magnitude. 
One may then ask whether this reduction of constraint violation is superficial---merely an expected outcome of increasing resolution (not yielding a true improvement in accuracy per cpu hour spent). 


Recall that Dendro uses an octree grid, subdividing each cell in half with each division. 
Thus the grid resolution $dx$ of a given cell relates to the mesh size $\mu$ 
(the number of mesh elements or cells in the grid) as $dx \propto \mu^{-1/3}$. 
We currently use 6th order spatial derivatives, so error decreases as $\varepsilon \propto dx^6$. Globally refining the grid would thus decrease errors as $\varepsilon \propto \mu^{-2}$; we would thus expect $\varepsilon \, \mu^2$ to remain constant. 
As WAMR refines the grid locally rather than globally, refining an unimportant region of the grid will do less to reduce global errors than refining an important region of the grid. Therefore, lower values of 
$\varepsilon \, \mu^2$ between runs implies improved refinement schemes: 
not merely that we have increased refinement, but that we have done so efficiently, allocating resources where they are most needed. 
To this end, we define a notion of ``grid efficiency'' as 
\begin{equation} \label{eqn:eta}
  \eta \equiv {C_{\rm tot}}^{-1} \mu^{-2}
\end{equation}
where $C_{\rm tot}$ is the $L_2$ norm of both Hamiltonian and momentum constraint violations, $\cal H$ and $\mathcal{M}_i$, across the entire grid and $\mu$ is again the mesh size. 
Comparing $\eta$ thus measures the relative efficiency of grid subdivision schemes: 
do our choices for refinement result in appropriate error reduction, 
or do they show some relative (in)efficiency? 

The lower panel of Figure~\ref{fig:CM} compares grid efficiency $\eta$ between v2022 and v2025. 
Not only do we reduce constraint violations by one to two orders of magnitude in the upgrade, we see similar scale improvements in grid refinement efficiency, 
accounting for the expected accuracy differences due to refinement. 
This shows that these gains are not due to superficial changes in refinement. 

While pre-merger refinement primarily serves to prevent errors (as WAMR does explicitly), post-merger refinement called for by ORBIT serves to propagate GWs out to the detection radii. 
For this reason, we expect the loss of efficiency in the ORBIT-enabled run, particularly as the high-frequency ringdown (here at $t/M \sim 625$) propagates out to the most distant GW detection radius ($100\,M$). 
The slight inefficiency incurred by enabling ORBIT thus points to the difference in \textit{purpose} of ORBIT refinement compared to WAMR refinement.

\subsection{Code updates accelerate runs} \label{sec:run_times}

\begin{figure}\centering
  \includegraphics[width=\linewidth]{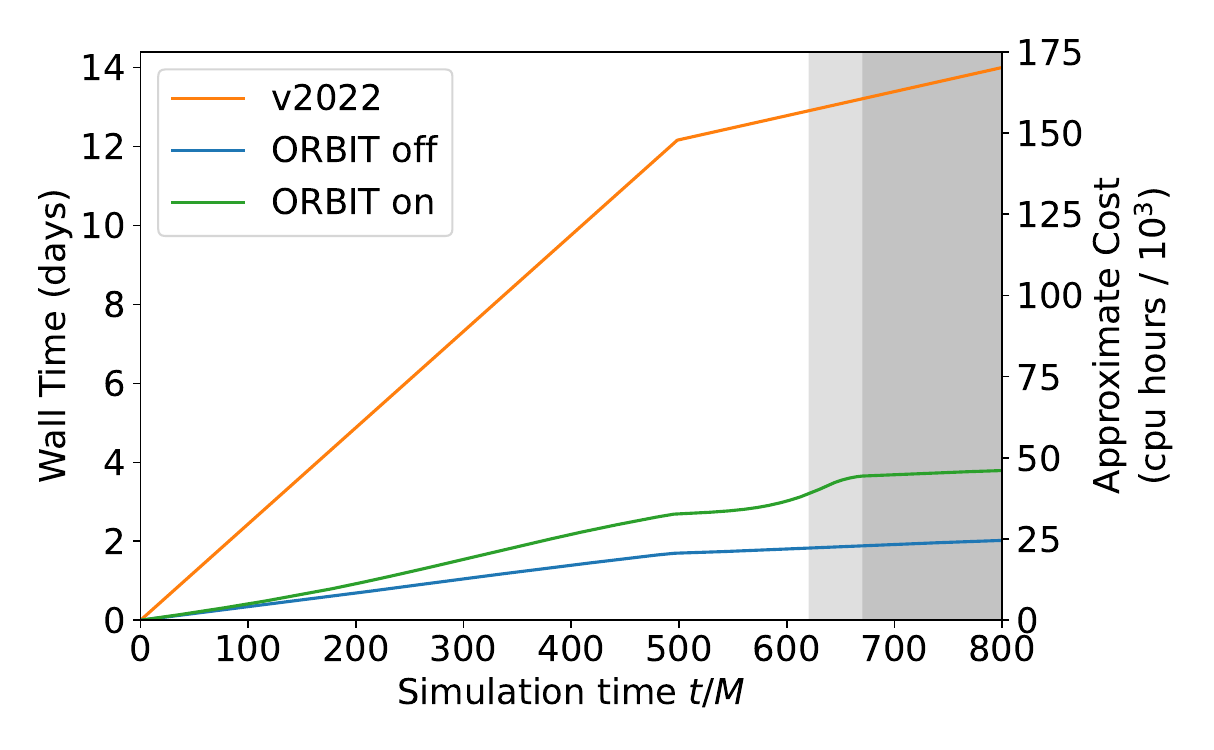}
  \vspace{-.75cm}
  \caption{
    Comparison of $q=4$ run cost between code versions, both in wall time as well as in approximate cpu hours. 
    Both humans and computers see four times lower run cost in v2025 than in v2022. 
    Coloring and shading as in Figure~\ref{fig:CM}. 
  }
  \label{fig:time}
\end{figure}

Recent changes to our algorithm achieve dramatic improvements in run cost---both in human time (wall hours) and in computer time (cpu hours). 
Figure~\ref{fig:time} shows that while v2022 took over two weeks to run, a current v2025 run takes less than four \textit{days}, cutting down wall time by a factor of $3.7$. 
Furthermore, computational cost is cut down by a factor of almost $4$. 
(This is a non-trivial distinction, as hypothetically the wall time could have been decreased by submitting identical code to four times the number of processors.) 

The ``ORBIT on'' run shows a slight bump in wall time around $625\,M$, as ORBIT propagates out to sustain the high-frequency ringdown. 
As illustrated in Figures~\ref{fig:progress_grid} \&~\ref{fig:omega-lambda}, post-merger refinement is the most costly, as it fills a 3D sphere with high refinement. 
While enabling ORBIT makes the run almost twice as expensive as disabling it, we find that the resulting gains in waveform quality (see e.g. Figure~\ref{fig:Psi4_l6m6_cf}) justify the cost.

\subsection*{}
\vspace{-1cm}

Updates to the code thus not only improve waveform quality and reduce constraint violations, but they do so efficiently and with decreased net computational cost.

\section{Conclusions} \label{sec:conclusions}

We introduced ORBIT, an optimized resolution baseline derived from BH inspiral trajectories, and showcase recent advances in Dendro-GR over the past few years. 
These improvements work toward fulfilling the need of next-generation GW detectors for higher accuracy in NR simulations. 

ORBIT causally refines Dendro's grid to sustain propagation of high-order spherical harmonic modes. 
The Nyquist criterion---combined with the relationship between BH orbital frequency $\omega_{\rm orbit} = 2 \pi f_{\rm orbit}$ and GW frequency---gives a resolution floor of 
\begin{equation}
  \Delta x \leq c / (2 \, m \, f_{\rm orbit}) \leq \lambda_{\rm Nyquist}
\end{equation}
as a sufficient criteria for Nyquist-resolving GW spin-weighted spherical harmonics of order $m$. 
ORBIT captures low-amplitude features otherwise missed by error-driven refinement methods like WAMR. 
As compared to maintaining high resolution across the entire duration of the simulation, this dramatically reduces computational cost while still maintaining necessary resolution. 

This 2025 version of Dendro-GR, with ORBIT and other improvements, demonstrates significant progress: 
\begin{itemize}
  \item Waveform noise has decreased by one to two orders of magnitude. 
  \item Constraint violations are reduced by two to three orders of magnitude. 
  \item Refinement efficiency has improved by two orders of magnitude. 
  \item Wall time and total run cost is reduced by a factor of four. 
\end{itemize}
In addition to ORBIT, other improvements include implementing Slow-Start Lapse (SSL), our custom Hamiltonian Damping scheme (HD), causal WAMR, and central grid structure updates (see \S\ref{sec:Dendro} for details on each of these changes). 
By reducing the sharpness of the initial lapse wave, SSL relaxes grid refinement requirements and reduces noise on the grid. 
Though the constant remeshing required as BHs travel across the grid generates Hamiltonian constraint violation $\cal H$ (as grid patches interpolate to different resolutions), 
our HD scheme damps total $\cal H$ by an order of magnitude, preventing its accumulation during inspiral. 
Our new causal WAMR structure puts refinement where we need it by ignoring spacelike regions of the grid (causally disconnected to the BH evolution) and by avoiding regions contaminated by the (dampened) lapse wave. 
In total, these improvements allow Dendro-GR to run at about one-fourth the computational cost while reducing constraint violations by orders of magnitude. 
These gains position Dendro-GR as an increasingly capable tool for simulating binary BH mergers, particularly at higher mass ratios. 

We anticipate additional improvements to Dendro-GR. 
For these first pass, proof-of-concept results, our implementation of ORBIT was hardcoded (see Appendix~\ref{apx:ORBIT}); we plan a softcoded implementation for greater flexibility. Additionally, we expect a need to extend ORBIT further as we test it more thoroughly with systems having high spins or large eccentricities.  These will certainly present new challenges. 

In conclusion, ORBIT and other recent improvements to Dendro-GR reduce waveform noise while lowering computational costs. These and other planned advances bring us closer to meeting the demands of next-generation GW detectors and to expanding our understanding of BBH systems across a wider range of systems.

\section*{Acknowledgments} 

This research was supported by NSF grants PHY-2207615 (BYU) and 
PHY-2207616 (Utah), as well as NASA grant 80NSSC20K0528. 
WKB is supported by the College of 
Computational, Mathematical, and Physical Sciences at BYU.  
Supercomputer time was provided by
\href{https://access-ci.org}{ACCESS} on \href{https://tacc.utexas.edu/systems/stampede3/}{Stampede3}.

We thank our collaborators Yosef Zlochower and Chloe Malinowski 
for many discussions during this work. 
WKB thanks JK Black for assistance in formulating Eq.~\ref{eqn:ell_N}
and thanks ZB Etienne for conversations on constraint damping. 
%

%

\section*{Data Availability} \label{sec:data_avail}

The version of {\sc Dendro-GR} used to produce the results 
in this paper is distributed with the MIT license at 
\href{https://github.com/paralab/Dendro-GR}{github.com/paralab/Dendro-GR}. 

Parameter files are on Pastebin for the ORBIT-enabled runs ($m_{\max} = 12$) 
  for $q=1$ (\href{https://pastebin.com/q9715xvz}{q9715xvz}) 
  and $q=4$ (\href{https://pastebin.com/Zxsn486e}{Zxsn486e}). 
The parameter \verb`BSSN_NYQUIST_M = 12` in these files requires ORBIT to resolve up to $m_{\max} = 12$ (see Eq.~\ref{eqn:dx_ORBIT}); the two matching $m_{\max} = 0$ runs change only this parameter (but have otherwise identical parameter files). 
Each run begins with the BHs along the $x$-axis at coordinate separation $x_2 - x_1 = 8\,M$ (with the center of mass at the origin), without spin, with $+ \hat z$ momenta giving a semi-circular orbit in the $xy$-plane. 

\href{https://doi.org/10.5281/zenodo.14649176}{Zenodo} hosts output data for this paper's runs \citep{Zenodo_Black_2025}. 
Plots visualized using \href{https://bitbucket.org/wkblack/dendro-gr-analysis-tools/src/master/}{wkblack/dendro-gr-analysis-tools} on Bitbucket. 
Some video visualizations are available on You\-Tube for the ORBIT-enabled runs; see Table~\ref{tab:videos}. 

\begin{table}[!ht]\centering
  \begin{tabular}{cccccc}
    \hline
    Run & Cell Level & Lapse & Waveform & Constraints & 3D \\
    \hline
    $q = 1$ 
    & \href{https://youtu.be/ONaAsLtaYYg}{far}, \href{https://youtu.be/HedInQbbajI}{near} 
    & \href{https://youtu.be/CHIMsXXgs24}{far} 
    & \href{https://youtu.be/GITZM5X5oGA}{$|\Psi_4|$}, \href{https://youtu.be/PaN5UHLI4lc}{$\mathfrak{Re}[\Psi_4]$}
    & \href{https://youtu.be/uH71kSTF-kw}{far}, \href{https://youtu.be/HTr-ScQIGVE}{near} 
    & \href{https://youtu.be/y8s17HSWWxU}{link} \\
    $q = 4$ 
    & \href{https://youtu.be/pbo_CZDLgQM}{far}, \href{https://youtu.be/Es1MMRUH0oQ}{near} 
    & 
    \href{https://youtu.be/jxVGSb0rs7s}{far} 
    & \href{https://youtu.be/l26wjAX77sU}{$|\Psi_4|$}, \href{https://youtu.be/iI06i4zcU1M}{$\mathfrak{Re}[\Psi_4]$} 
    & \href{https://youtu.be/u7JEkLvmAso}{far}, \href{https://youtu.be/0DG7yKnjd5w}{near} 
    & \href{https://youtu.be/uzMQOTdY4jY}{link} \\ 
    \hline
  \end{tabular}
  \caption{
    ORBIT-enabled runs \href{https://www.youtube.com/playlist?list=PLCmfEjDzshzOP9rLkN0s9T-MtBk-fmAo6}{visualized} in various fields. 
  }
  \label{tab:videos}
\end{table}

%

\bibliographystyle{apsrev4-2}
\bibliography{main.bib}

\appendix

\section{Central refinement about the BHs} \label{apx:central_refinement}

Here we determine core refinement about each black hole. 
To ensure we have $N$ points across the horizon of a BH of radius $r_{\rm BH}$ on an octree grid of size $L$ with element order $n$ (determining the number of points across a cell), we must keep the level of refinement $\ell$ (where larger $\ell$ indicate higher resolution) about the BH to be at least
\begin{equation} \label{eqn:ell_N}
  \ell \geq 2 + \log_2\left( \frac{L}{n \cdot R_{\rm BH}} \left\lfloor \frac{N + 1}{2} \right\rfloor \right) 
\end{equation}
(where $\lfloor * \rfloor$ is the floor operator). 
In our evolution coordinates, $R_i = m_i$ \footnote{
  The initial coordinates have $R_{i} = m_i / 2$, but this quickly settles (after $t/M < 10$) into the evolution coordinates of $R_i = m_i$. 
}; the larger BH has a mass of $m_1 = M / (1 + q)$ while the smaller BH has a mass of $m_2 = M / (1 + 1/q)$. 
Additionally accounting for our box size $L = 800\,M$, element order $n = 6$, and our goal of $N \geq 50$ gives us level requirements about each BH of 
\begin{align}
  \ell_1 &\geq 13.7 + \log_2 ( 1 + q ) \\
  \ell_2 &\geq 13.7 + \log_2 ( 1 + 1/q )
\end{align}
(where $m_{\rm BH, 1} \leq m_{\rm BH, 2}$). 
These then dictate $\ell_{{\rm BH}, i} \geq 15$ for $q=1$ while for $q=4$ it dictates 14 and 16 for the larger and smaller BHs respectively. More generally, as we approach extreme mass ratios, we find we need cell level 14 and $14 + \log_2 q$ as $q \rightarrow \infty$ given our grid size and element order. 
We note there exists a two-level offset in the cell level parameters $\ell_i$ from the resulting level on the grid, such that the above equations enforce cell level 13 for $q=1$.

%

\section{More on ORBIT} \label{apx:ORBIT}

Here we present more details regarding our Nyquist-compliant refinement floor ORBIT.

\subsection{Choice of target order} \label{apx:error_Nyquist}

The Nyquist criterion dictates the minimum refinement necessary to capture and reconstruct a frequency; lower refinements will preclude propagation of the signal. 
Fractional error of a signal frequency $f_{\rm signal}$ as a function of sampling rate $f_{\rm sample}$ is given by
\begin{equation} \label{eqn:error_Nyquist} 
  \varepsilon = 1 - \frac{1}{\sqrt{1 + (f_{\rm signal} / f_{\rm sample})^{2}}},
\end{equation}
where $f_{\rm Nyquist} = 2 \, f_{\rm signal}$. 
At high sampling rate (in the regime where $f_{\rm sample} \gg f_{\rm signal}$), this error approaches $\varepsilon = \frac12 (f_{\rm signal} / f_{\rm sample})^2$: beyond $f_{\rm Nyquist}$, error falls off as sampling frequency squared. 
Thus, if we Nyquist-resolve mode $m$ (i.e. $f_{\rm sample} = 2 f_{\rm GW}^{(l,m)} = f_{\rm Nyquist}$), we have $\varepsilon \sim 10\%$. 
Nyquist-resolving to $2m$ ($f_{\rm sample} = 2 f_{\rm Nyquist}$) would then have only error $\varepsilon \sim 3\%$ on mode $m$ 
while Nyquist-resolving $7m$ would yield $\varepsilon \sim 1\%$ 
(lower-frequency waveforms would improve even more so). 
In this paper, we Nyquist-resolve $m=12$, decreasing aliasing-induced
errors in the target $(l,m) = (6,-6)$ waveform to $3\%$. 

\subsection{Calculation from orbital frequency}

\begin{figure}
    \centering
    \includegraphics[width=\linewidth]{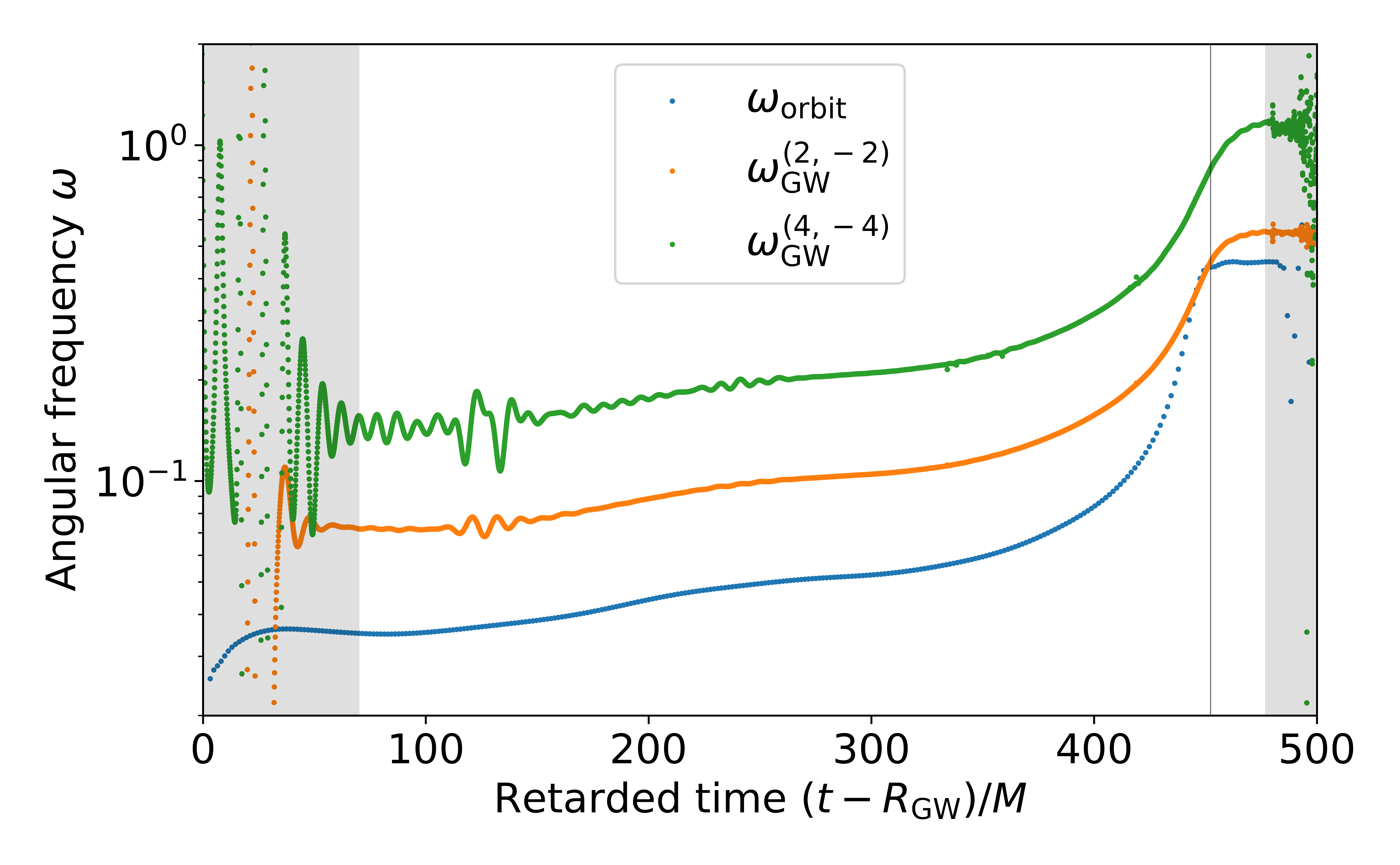}
    \includegraphics[width=\linewidth]{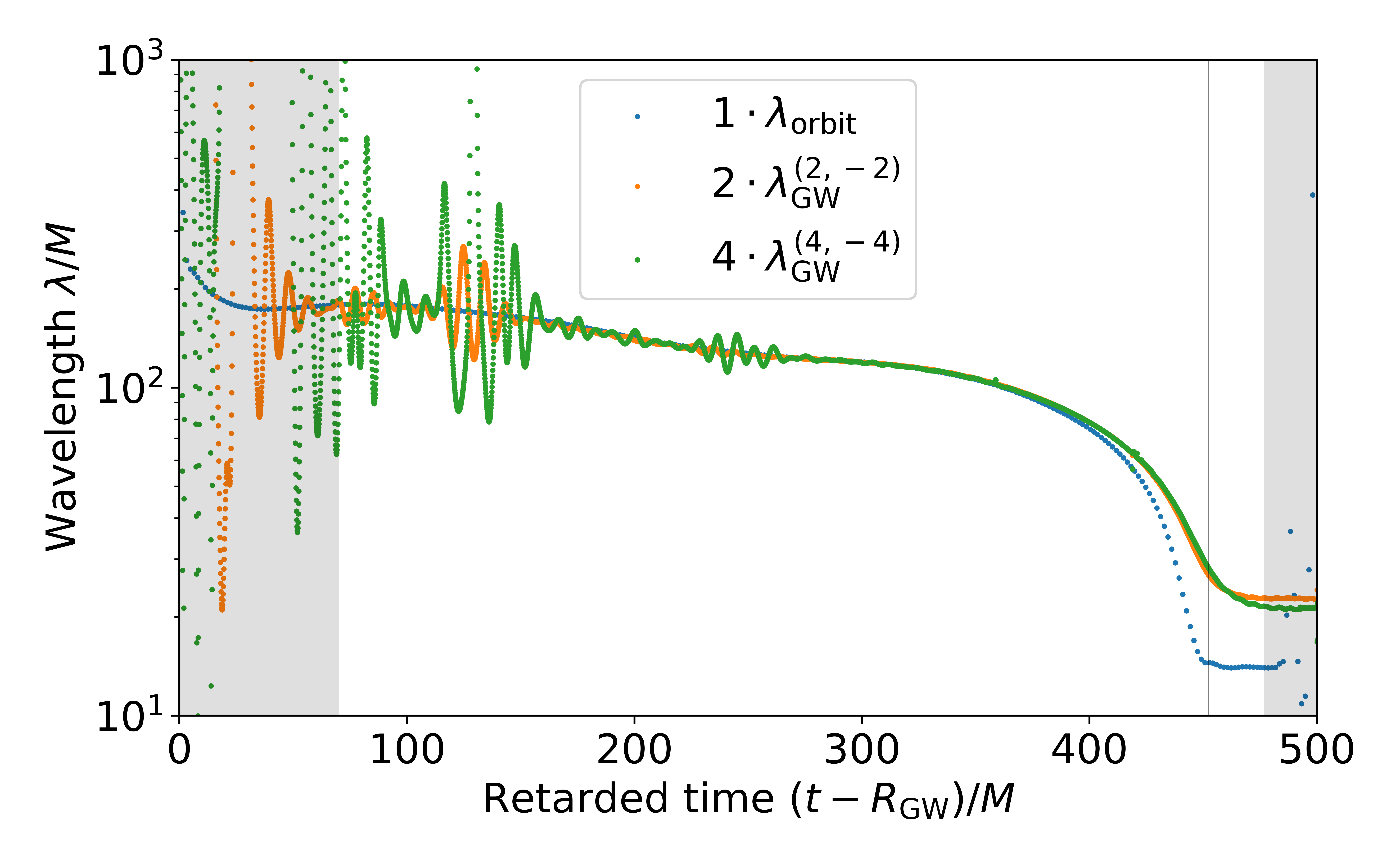}
    \vspace{-.75cm}
    \caption{
      Angular frequencies (upper plot) and corresponding rescaled wavelengths (lower plot) from a $q=1$ binary inspiral. 
      In blue are curves from BH coordinate positions; 
      the vertical line indicates the point where BH separation distance is $ds = 1\,M$ (near this time the BHs form a common horizon 
      and their coordinate positions decorrelate with GW emission). 
      In orange and green are data from the $(l, m) = (2, -2)$ and $(4, -4)$ GW modes. 
      Because $\lambda_{\rm orbit} \lesssim m \lambda_{\rm GW}^{(l,m)}$, 
      refining to $\Delta x = \lambda_{\rm orbit} / 2 |m|$ 
      is sufficient to Nyquist-resolve GWs of order $\leq |m|$. 
      Shaded regions as in Figure~\ref{fig:Psi4_qx}. 
      \vspace{-.5cm}
    }
    \label{fig:omega-lambda}
\end{figure}

We next discuss the relationship between BH orbital frequency and GW wavelength. 
The upper panel of figure~\ref{fig:omega-lambda} shows angular orbital frequencies $\omega$ for the BH orbit as well as for the emitted GWs; the lower panel shows the corresponding rescaled wavelengths $m \lambda$ 
(with orbital wavelength $\lambda_{\rm orbit} \equiv 2 \pi c / \omega_{\rm orbit}$) each overlapping. 
Angular frequencies for the BHs are calculated from the progression of 
the relative angle between the BHs in the orbital plane 
and for the GWs from the phase angle in complex space. 
The lower plot then rescales by $m$, showing universal agreement in waveform across $m$ in the early inspiral. 
As we approach merger, $\lambda_{\rm orbit} < m \lambda_{\rm GW}^{(l,m)}$, implying that BHs inspiral faster than their emitted wavelengths. 
At merger, the BHs form a common horizon and 
their coordinate locations decorrelate from the emitted GWs. 
This works to our favor, as refining to grid spacing 
$\Delta x = \lambda_{\rm orbit} / 2|m|$ 
gives a stricter refinement criterion than refining to 
$\Delta x = \lambda_{\rm GW}^{(l,m)} / 2$; 
using rescaled orbital wavelength therefore \textit{anticipates} refinement requirements, 
ensuring the grid already has sufficient resolution for the GWs 
before they reach a given portion of the grid. 
As wavelengths cannot always be known \textit{a priori}, 
this provides us with a sufficient resolution floor 
to satisfy the Nyquist criterion \textit{in situ}. 
ORBIT thus adapts to support the emitted GWs mid-simulation.

\subsection{Our formulation for cell level}

Given an input orbital wavelength, one can then determine the 
particular refinement level necessary to support GWs up to order $|m|$. 
Determining the target resolution level $\ell$ depends on several simulation-specific factors: 
  we use a 6th-order derivative scheme ($n = 6$) on a box size of $L = 800\,M$; Dendro resolution level $\ell$ gives the sub-division of the octree: $\ell = 0$ would be a single cube (low resolution) while higher levels repeatedly sub-divide that base cube by factors of two. 
Dividing the target wavelength $\lambda_{\rm orbit}$ by two for 
the Nyquist criterion and dividing by the order $m$ of a target mode 
$\lambda_{\rm GW}^{(l,m)}$ yields a minimum resolution level requirement of: 
\begin{equation} \label{eqn:ell_Nyquist_m}
  \ell_{\rm Nyquist}(m) 
  = \left\lceil \log_2 \frac{2L / n}{\lambda_{\rm GW}^{(l,m)}} \right\rceil
  \leq \left\lceil \log_2 \frac{800\,M / 3}{\lambda_{\rm orbit}/m} \right\rceil 
\end{equation}
(where $\lambda_{\rm GW}^{(l,m)}$ is the target wavelength, $\lambda_{\rm orbit} / m$ Nyquist-resolves mode $m$, and $\lceil * \rceil$ is the ceiling operator). 

\begin{table}\centering
  \setlength{\tabcolsep}{6pt}
  \begin{tabular}{ccccc}
    \hline
    Run & $A$ & $\tau_0$ & $\lambda_{\min}$ & $t_{\rm end}$ \\
    \hline
    $q = 1$ & 18.8 & 445.0 & 14.49 & 580.14 \\
    $q = 4$ & 17.0 & 490.0 & 18.10 & 670.40 \\
    \hline
  \end{tabular}
  \caption{
    Fit parameters for Eq.~\ref{eqn:lambda_fit}, modeling BH orbital wavelength $\lambda_{\rm orbit}$. 
    Amplitude of inspiral given by $A$, merger time estimated as $\tau_0$, minimum orbital wavelength given by $\lambda_{\min}$, and ORBIT end time given by $t_{\rm end}$ (see Eq.~\ref{eqn:causal_end}). 
  }
  \label{tab:lambda_params}
  \vspace{-.5cm}
\end{table}

While Dendro-GR now softcodes ORBIT (actively \textit{in situ} tracking BH orbital frequency), these first-pass proof-of-concept runs were prototyped with an \textit{a priori} hardcoded formulation (modeling BH orbital frequency from past high-quality runs).
Following quasi-circular orbit results, we used 
\begin{equation} \label{eqn:lambda_fit}
  \lambda_{\rm orbit} \approx 
  \begin{cases}
    \tau > 0 :& \max \left( A \, \tau^{3/8}, \, \lambda_{\min} \right) \\
    \tau < 0 :& \lambda_{\min}
  \end{cases}
\end{equation}
(where $\tau \equiv \tau_0 - t$ is time to merger); 
Table~\ref{tab:lambda_params} gives best fit parameters. 
Adding the first post-Newtonian term changes the formula to 
$\lambda = A \, \tau^{3/8} (1 + B \, \tau^{-2/8})$
but $B$ is consistent with zero for $q=1$ and for $q=4$ the fit works well enough with $B=0$ for most of the inspiral; near plunge the simpler fit outpaces the more complex fit, giving a slightly more conservative refinement criterion. 
Rather than adding further post-Newtonian expansion terms to improve the fit, we will soon have a version of Dendro-GR which tracks past BH orbital frequency mid-simulation, giving live updates to refinement for future, heretofore unexplored mass ratios.

\begin{figure*}\centering
  \includegraphics[width=.49\linewidth]{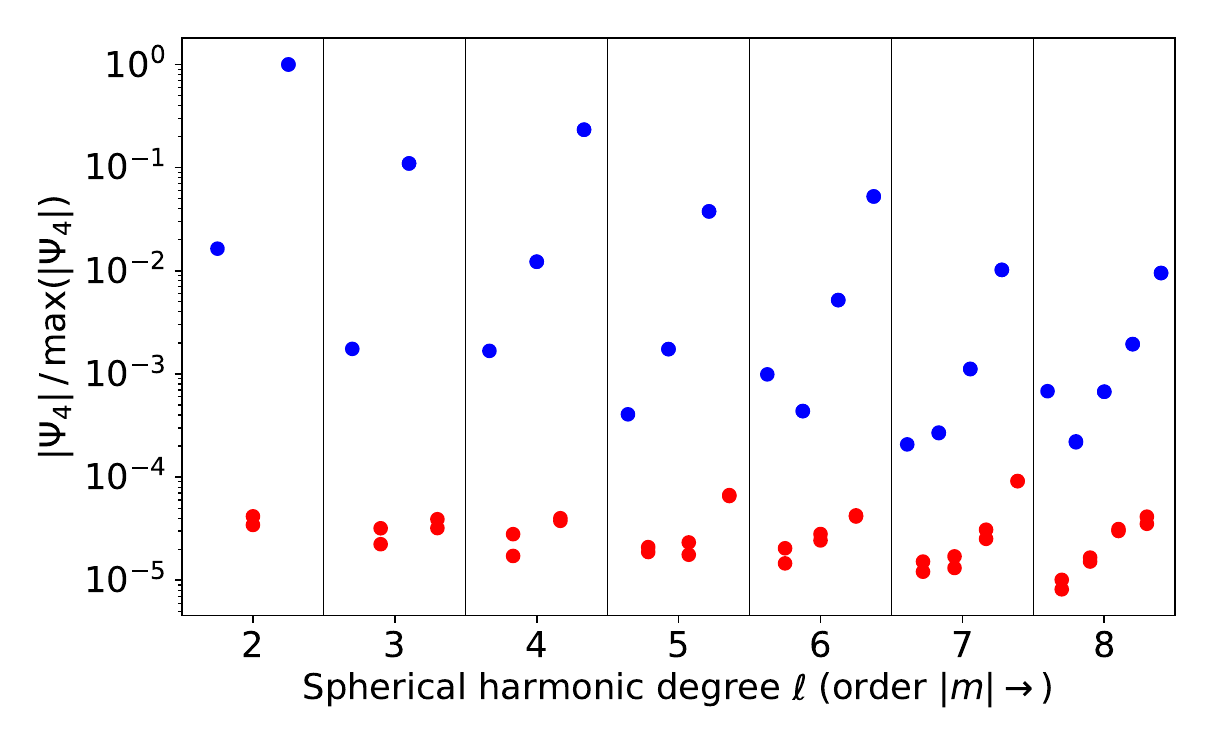}
  \includegraphics[width=.49\linewidth]{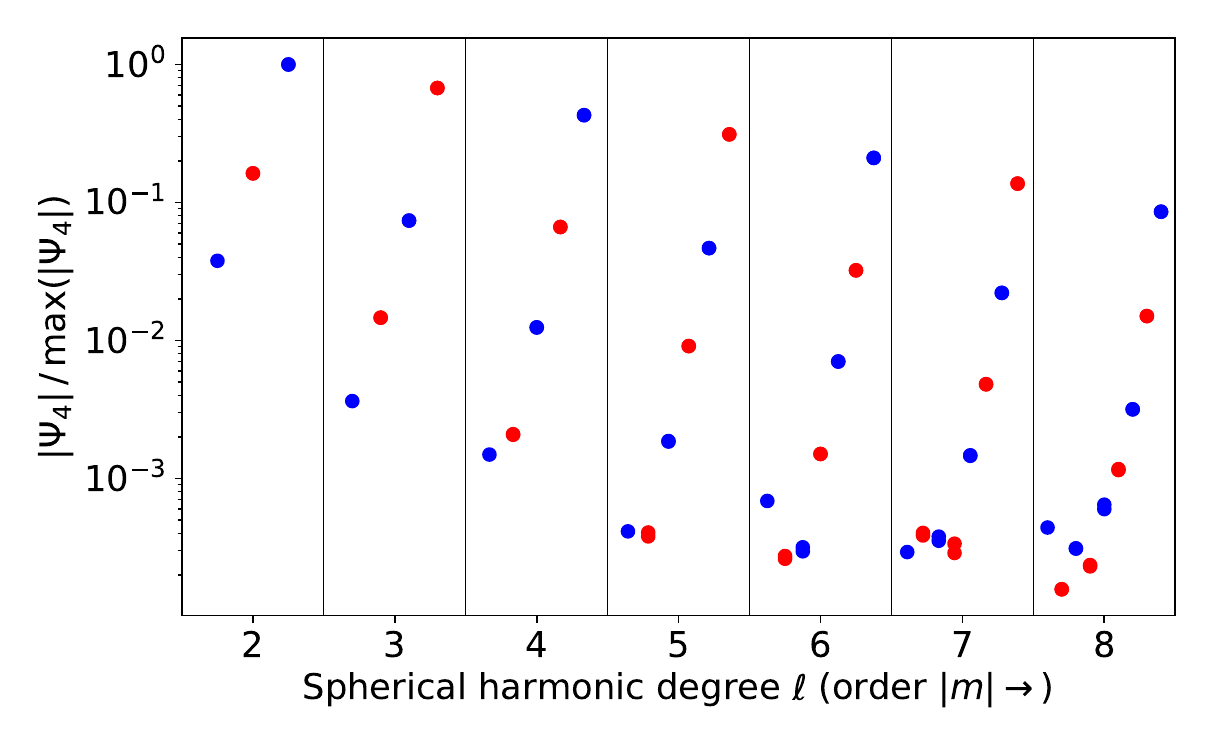}
  \vspace{-.25cm}
  \caption{
    Mode amplitudes at merger for $q=1$ (\href{https://media1.giphy.com/media/v1.Y2lkPTc5MGI3NjExb2l5NWtzYXZkbnJuMnAxd3I5NTdrYWY5eXVreWE3a2cwdHd3MzlyMCZlcD12MV9pbnRlcm5hbF9naWZfYnlfaWQmY3Q9Zw/eNv9bJES74Ve976eQu/giphy.gif}{left}) and $q=4$ (\href{https://media0.giphy.com/media/v1.Y2lkPTc5MGI3NjExOGY1YXl1N2hiNmpuMWhsYzUwb2dvMDhsdnh5c3RlYjcyYTlibGRyciZlcD12MV9pbnRlcm5hbF9naWZfYnlfaWQmY3Q9Zw/SUqwtjTeYHMsxVrHm2/giphy.gif}{right}). [See links in the parentheticals for evolution videos.] 
    Within each vertical cell, spherical harmonic order $|m|$ increases toward the right (with odd $m$ colored red for visibility). 
    The first cell thus displays mode amplitudes $(2,0)$, $(2,\pm 1)$, and $(2,\pm 2)$; the final cell displays $(8,0)$, $(8,\pm 1)$, \dots $(8, \pm 8)$. 
    Errors are small enough that the $+m$ and $-m$ modes lie on top of each other for sufficiently high amplitude modes. 
  }
  \label{fig:mode_strength}
\end{figure*}

\section{Mode coupling of waveforms at merger} \label{apx:merger_amp}

Mode coupling strengthens with mass ratio $q$: spherical harmonic modes lie closer together at higher mass ratios. The more extreme the mass ratio, the more nonlinear effects plague the simulation. 
We can quantify this mode coupling by observing how the spin-weighted spherical harmonic amplitude at merger $|\Psi_4^{(l,m)}|_{\max}$ decreases with degree $l$ (it \textit{increases} with order $|m|$). 
The ratio of modes follows steady patterns, 
where $|\Psi_4^{(l,l)}|_{\max} / |\Psi_4^{(l+1,l+1)}|_{\max}$ is roughly constant 
(as well as $|\Psi_4^{(l,m)}|_{\max} / |\Psi_4^{(l,m+1)}|_{\max}$). 
We can quantify this trend across many mass ratios, measuring the log slope of $|\Psi_4^{(l,m)}|_{\max}$ as a function of $l$, $m$, and $q$. 
This allows extrapolation of high-frequency modes, avoiding the expense of direct calculation.

Figure~\ref{fig:mode_strength} shows relative GW amplitudes at merger (normalized at the maximum amplitude) for both $q=1$ and $q=4$. The left and right plots have different vertical scales: most of the modes for $q=4$ are of higher relative amplitude than those of $q=1$; they therefore have far smaller errors (see also Figure~\ref{fig:Psi4_qx}). 

There are a few ways to estimate error on these values. 
The symmetries of our $q=1$ run imply that all odd $m$ modes (colored red) should have null amplitude; their non-zero values thus roughly estimate epistemic error. 
Mirroring $\pm m$ mode amplitudes should have equal amplitude; differences between values thus roughly estimates aleatoric error. 
Accounting for these uncertainties, clear trends emerge across $l$, $m$, and $q$. 

Fitting $l = |m|$ (excluding odd $m$ for $q=1$) amplitudes for non-spinning, non-elliptical mergers, we find a steady flattening of log slope as a function of symmetric mass ratio $\nu \equiv q / (1 + q)^2$. 
This shows that higher-degree modes are increasingly important for unequal mass ratios. 
Fitting the results of $q \in \{1, 2, 4, 8\}$, we find a clean power law relation of slope versus symmetric mass ratio: 
\begin{equation}
  \frac{d}{dl} \log_{10} |\Psi_4^{(l,l)}|_{\max} = -{\rm e}^{(7.23 \pm 0.07) \nu - (2.91 \pm 0.01)} 
\end{equation}
Thus we have relative mode strength (quantified as the magnitude of the power law slope of the spherical harmonic decomposition main modes) 
increasing in amplitude toward more equal mass ratios ($\nu \rightarrow 1/4$) and decreasing in amplitude toward more extreme mass ratios ($\nu \rightarrow 0$). 
While at $q=1$, an increment in $l$ results in roughly a factor of two loss of relative amplitude, 
as $q\rightarrow \infty$, incrementing $l$ results in only a $\sim 10\%$ loss of amplitude. 
While for $q=1$ the $(l,m)=(2,\pm 2)$ mode makes up $\lesssim 1/2$ of the total signal, the same mode for $q \rightarrow \infty$ only makes up $\lesssim 10\%$ of the total signal. 
This then quantifies the relative importance of higher-amplitude modes at more extreme mass ratios at merger. 


Not only do we see clear trends in $|\Psi_4^{(l,l)}|_{\max}$; we also see rich structure within lower-$m$ modes. 
The right panel of Figure~\ref{fig:mode_strength} (with $q=4$) has larger amplitude modes and is thus less contaminated by noise; this better resolves the non-dominant modes, revealing an inner structure to the mode amplitudes. 
The patterns in $(l,m)$ mode amplitudes hint at the ability to extrapolate to infinite degree (traveling out on the top bar of $l=|m|$, then filling in the inner structure with decreasing $|m|$), given only the trends from the first few $(l,m)$ modes. This would incur relatively small extrapolation error, as modes grow progressively weaker with increased extrapolation. 
As estimation of mode amplitudes grows increasingly expensive (and error-fraught) toward high $l$, extrapolating higher modes would decrease computational expense. 
This may well produce more accurate waveforms than using the raw measured waveform amplitudes (which are infiltrated by noise around $10^{-3}$ relative amplitude). 

More generally, these trends could perhaps even be extrapolated out toward more extreme mass ratios, spins, and eccentricities, moving toward a universal fit to waveform amplitudes at merger.

%


%


\end{document}